\documentclass[conference, 10pt]{IEEEtran}
\IEEEoverridecommandlockouts
\usepackage{cite}
\usepackage{amsmath,amssymb,amsfonts}
\usepackage{algpseudocode}
\usepackage{graphicx}
\usepackage{textcomp}
\usepackage{xcolor}
\usepackage{braket}
\usepackage{url}
\usepackage{hyperref}
\usepackage{booktabs} 
\usepackage{multirow} 
\usepackage{makecell}

\hypersetup{
  colorlinks   = true, 
  urlcolor     = blue, 
  linkcolor    = blue, 
  citecolor   = blue 
}

\renewcommand{\figureautorefname}{Figure~\negthinspace}

\renewcommand{\tableautorefname}{Table~\negthinspace}

\def\BibTeX{{\rm B\kern-.05em{\sc i\kern-.025em b}\kern-.08em
    T\kern-.1667em\lower.7ex\hbox{E}\kern-.125emX}}
\begin{document}

\title{Differentiable Quantum Architecture Search in Quantum-Enhanced Neural Network Parameter Generation \thanks{The views expressed in this article are those of the authors and do not represent the views of Wells Fargo. This article is for informational purposes only. Nothing contained in this article should be construed as investment advice. Wells Fargo makes no express or implied warranties and expressly disclaims all legal, tax, and accounting implications related to this article.}
}


\author{
\IEEEauthorblockN{ 
    Samuel Yen-Chi Chen\IEEEauthorrefmark{1} \IEEEauthorrefmark{8},
    Chen-Yu Liu \IEEEauthorrefmark{2}\IEEEauthorrefmark{11},
    Kuan-Cheng Chen\IEEEauthorrefmark{3}\IEEEauthorrefmark{4}\IEEEauthorrefmark{12},\\
    Wei-Jia Huang\IEEEauthorrefmark{5}\IEEEauthorrefmark{13}, 
    Yen-Jui Chang\IEEEauthorrefmark{6}
    \IEEEauthorrefmark{7} \IEEEauthorrefmark{14},
    Wei-Hao Huang\IEEEauthorrefmark{9}\IEEEauthorrefmark{10}
}

\IEEEauthorblockA{\IEEEauthorrefmark{1}Wells Fargo, New York, NY, USA}
\IEEEauthorblockA{\IEEEauthorrefmark{2}Graduate Institute of Applied Physics, National Taiwan University, Taipei, Taiwan}
\IEEEauthorblockA{\IEEEauthorrefmark{3}Department of Electrical and Electronic Engineering, Imperial College London, London, UK}
\IEEEauthorblockA{\IEEEauthorrefmark{4}Centre for Quantum Engineering, Science and Technology (QuEST), Imperial College London, London, UK}
\IEEEauthorblockA{\IEEEauthorrefmark{5}Hon Hai (Foxconn) Research Institute, Taipei, Taiwan}
\IEEEauthorblockA{\IEEEauthorrefmark{6}Quantum Information Center, Chung Yuan Christian University, Taoyuan City, Taiwan}
\IEEEauthorblockA{\IEEEauthorrefmark{7}Master Program in Intelligent Computing and Big Data, Chung Yuan Christian University, Taoyuan City, Taiwan}
\IEEEauthorblockA{\IEEEauthorrefmark{9}Jij inc., Tokyo, Japan}

\IEEEauthorblockA{Email:\IEEEauthorrefmark{8} \texttt{yen-chi.chen@wellsfargo.com}, \IEEEauthorrefmark{10} \texttt{w.huang@j-ij.com}, \IEEEauthorrefmark{11} \texttt{d10245003@g.ntu.edu.tw}, \\ \IEEEauthorrefmark{12} \texttt{kuan-cheng.chen17@imperial.ac.uk}, \IEEEauthorrefmark{13} \texttt{wei-jia.huang@foxconn.com}, \IEEEauthorrefmark{14}\texttt{aceest@cycu.edu.tw}}
}

\maketitle
\begin{abstract}
The rapid advancements in quantum computing (QC) and machine learning (ML) have led to the emergence of quantum machine learning (QML), which integrates the strengths of both fields. Among QML approaches, variational quantum circuits (VQCs), also known as quantum neural networks (QNNs), have shown promise both empirically and theoretically. However, their broader adoption is hindered by reliance on quantum hardware during inference. Hardware imperfections and limited access to quantum devices pose practical challenges. To address this, the Quantum-Train (QT) framework leverages the exponential scaling of quantum amplitudes to generate classical neural network parameters, enabling inference without quantum hardware and achieving significant parameter compression. Yet, designing effective quantum circuit architectures for such quantum-enhanced neural programmers remains non-trivial and often requires expertise in quantum information science. In this paper, we propose an automated solution using differentiable optimization. Our method jointly optimizes both conventional circuit parameters and architectural parameters in an end-to-end manner via automatic differentiation. We evaluate the proposed framework on classification, time-series prediction, and reinforcement learning tasks. Simulation results show that our method matches or outperforms manually designed QNN architectures. This work offers a scalable and automated pathway for designing QNNs that can generate classical neural network parameters across diverse applications.
\end{abstract}


\section{Introduction}
Quantum computing (QC) and quantum machine learning (QML) are rapidly advancing frontiers that promise to reshape computational approaches across a wide range of fields \cite{biamonte2017quantum, dunjko2016quantum}. Among various QML strategies, variational quantum algorithms (VQAs) \cite{cerezo2021variational, bharti2022noisy} have emerged as a prominent framework designed to operate on noisy intermediate-scale quantum (NISQ) hardware. Recent theoretical analyses have highlighted potential advantages of VQAs over classical machine learning under specific conditions, such as in terms of model expressivity and generalization behavior \cite{abbas2021power, du2020expressive, caro2022generalization,chen2024validating}. These capabilities have stimulated a growing body of work applying QML techniques to diverse tasks, including classification \cite{mitarai2018quantum, chen2021end, chen2022quantumCNN, qmlapp2, chen2024compressedmediq, chen2024qeegnet, lin2024quantumGradCAM, tseng2025transfer,qi2023qtn}, reinforcement learning \cite{chen2020variational, chen2022variational, yun2023quantum, chen2024efficient, lockwood2020reinforcement, skolik2022quantum, jerbi2021parametrized, chen2023QuantumDeepRecurrentRL, chen2023asynchronous, chen2023quantumDPER, chen2024learning}, time-series modeling \cite{chen2022quantumLSTM, chen2022reservoir, lin2024quantum, chen2024learning, chehimi2024federated}, and natural language processing \cite{yang2021decentralizing, yang2022bert, di2022dawn, stein2023applying, li2023pqlm}.

Despite recent progress, conventional QML approaches continue to face major obstacles—most notably, the difficulty of encoding large-scale classical data into quantum circuits. Common strategies such as gate-angle encoding and amplitude encoding are fundamentally constrained by the limited number of available qubits and the short coherence times of near-term quantum hardware, posing critical barriers to scalability and real-world deployment. In addition, the inference stage of QML models typically relies on access to quantum processors—often through cloud-based services—where hybrid quantum-classical computations are executed sequentially across circuit layers. This architectural dependence introduces latency and computational inefficiencies, especially in time-critical scenarios such as autonomous control or online decision-making.

To address these limitations, a hybrid architecture known as Quantum-Train (QT) has been proposed \cite{liu2024training, liu2024quantum, liu2024qtrl, lin2024quantum, liu2024federated, liu2024quantum2, lin2024quantum2, liu2024quantum3, liu2025federatedQTLSTM, chen2024QT_Dist_RL, liu2024QT_FWP, liu2024introduction, liu2025frame}. The core idea behind QT is to shift the role of quantum computation from data processing to parameter generation. Specifically, a quantum neural network (QNN) is trained to output the weights of a classical neural network, allowing classical data to be processed entirely on a classical architecture. This design eliminates the need for quantum data encoding and removes dependency on quantum hardware during inference, resulting in a fully classical model after training—well-suited for deployment in near-term applications.

However, existing QT frameworks still rely on manually crafted QNN architectures, often requiring expert-level domain knowledge and trial-and-error validation. This design bottleneck presents a steep barrier for widespread adoption outside the quantum machine learning community. To overcome this challenge, we propose Quantum-Train with Differentiable Quantum Architecture Search (DiffQAS-QT)-a novel framework that automates the discovery of high-performing QNN architectures specifically designed to generate classical neural network parameters for downstream learning tasks.

In this study, we demonstrate the versatility and effectiveness of DiffQAS-QT across a diverse set of machine learning tasks, including image classification, time-series forecasting, and reinforcement learning. Our experiments show that DiffQAS-QT delivers strong performance on MNIST classification, accurate rollout prediction in QT-enhanced LSTM models for temporal data, and robust policy learning in A3C-based QT-RL agents navigating MiniGrid environments. Notably, the proposed framework achieves not only high predictive accuracy but also superior training stability across modalities. These results suggest that DiffQAS-QT offers a promising direction for automating QNN design and advancing hybrid quantum-classical learning in practical settings.

\section{Related Works}
\label{sec:related_works}
Quantum-Train (QT) is a hybrid framework in which a quantum neural network (QNN) is employed to generate the parameters—such as weights and biases—of a classical neural network \cite{liu2024training, liu2024quantum, liu2024introduction}. By shifting the role of quantum computation from data processing to parameter synthesis, QT provides an alternative approach to hybrid learning that avoids the costly data encoding bottleneck common in many QML systems.

Since its initial formulation, the QT framework has been extended to a wide range of machine learning domains. In image classification, QT-based architectures have been applied to build compact yet expressive visual models \cite{liu2024quantum, liu2024quantum2, lin2024quantum2}. For time-series forecasting, QT-enhanced recurrent networks (QT-LSTM) and fast weight programmers have been explored for modeling temporal dynamics in both centralized and federated settings \cite{lin2024quantum, liu2025federatedQTLSTM, liu2024QT_FWP}. The framework has also been adopted for federated learning scenarios to provide privacy-preserving features \cite{liu2024federated, liu2025federatedQTLSTM}, enabling decentralized training of quantum-informed models without direct sharing of raw data. In addition, recent works have applied QT to natural language processing tasks \cite{liu2024quantum3, liu2025frame}, as well as to reinforcement learning, where QT-generated policies are trained within quantum-classical agent architectures \cite{liu2024qtrl, chen2024QT_Dist_RL}.
While numerous QML architectures such the QNN used in QT have demonstrated promising empirical performance, a major barrier to their widespread adoption in practical applications lies in the significant expertise required to design high-performing quantum circuits. In particular, constructing effective QNN architectures often demands deep domain knowledge in quantum information science, limiting accessibility to a broader machine learning audience. \emph{Quantum architecture search} (QAS) has emerged as a promising research direction aimed at alleviating this bottleneck. The goal of QAS is to automatically generate efficient quantum circuit architectures for target tasks—such as quantum state preparation or QML model training—with minimal or no manual intervention from the user \cite{martyniuk2024quantum}. A range of methods have been proposed to tackle this challenge. One popular line of work frames circuit construction as a sequential decision-making process, making reinforcement learning (RL) a natural candidate for QAS \cite{kuo2021quantum, ye2021quantum, chen2023QRL_QAS, dai2024quantum}. While RL-based approaches are theoretically universal solvers, they face several practical challenges, including the exploration–exploitation trade-off, sensitivity to optimizer hyperparameters, and high sample complexity. To address some of these issues, evolutionary QAS methods have been introduced \cite{ding2022evolutionary, chen2024evolutionary}. These methods encode quantum circuits into genetic representations (e.g., chromosomes), and evolve a population $\mathcal{P}$ of architectures through selection, crossover, and mutation over successive generations. Fitness scores—based on fidelity, QML task performance, or information-theoretic measures—guide the evolutionary process. A notable advantage of evolutionary QAS is its natural amenability to parallelization, allowing efficient deployment on high-performance computing (HPC) resources, particularly for batch evaluation of candidate circuits. Despite these advances, a fundamental limitation remains: both RL-based and evolutionary approaches operate in discrete search spaces, where the combinatorial explosion of candidate architectures can become prohibitive as the search space expands. Inspired by classical neural architecture search (NAS) \cite{liu2018darts}, recent works propose differentiable QAS methods \cite{ zhang2022differentiable, sun2023differentiable, chen2024differentiable, chen2025learning_to_measure}, where the architecture space is relaxed into a continuous domain using structural weights or parameterized distributions over candidate circuits. This relaxation enables gradient-based optimization over architecture parameters alongside circuit parameters in an end-to-end fashion.

In this work, we integrate differentiable QAS with the Quantum-Train (QT) framework. This combination enables us to explore a much larger architectural search space without suffering from the discrete sampling inefficiencies present in RL and evolutionary methods. Furthermore, it allows us to jointly optimize both the variational parameters within the quantum circuit and the architecture parameters controlling circuit structure, under a unified differentiable pipeline—ultimately enabling smooth, gradient-based training of quantum-generated classical models.
\section{Quantum Neural Networks}
\emph{Variational Quantum Circuits} (VQCs), also referred to as \emph{parameterized quantum circuits} (PQCs), represent a widely studied class of quantum models characterized by their tunable gate parameters. In modern hybrid quantum–classical learning paradigms, VQCs serve as the foundational building blocks for quantum neural networks (QNNs) \cite{bharti2022noisy, mitarai2018quantum}. A growing body of theoretical work has shown that VQCs are capable of exhibiting specific forms of quantum advantage under appropriate conditions \cite{abbas2021power, caro2022generalization, du2020expressive}, underscoring their relevance to quantum machine learning applications.
\begin{figure}[htbp]
\begin{center}
\includegraphics[width=1\columnwidth]{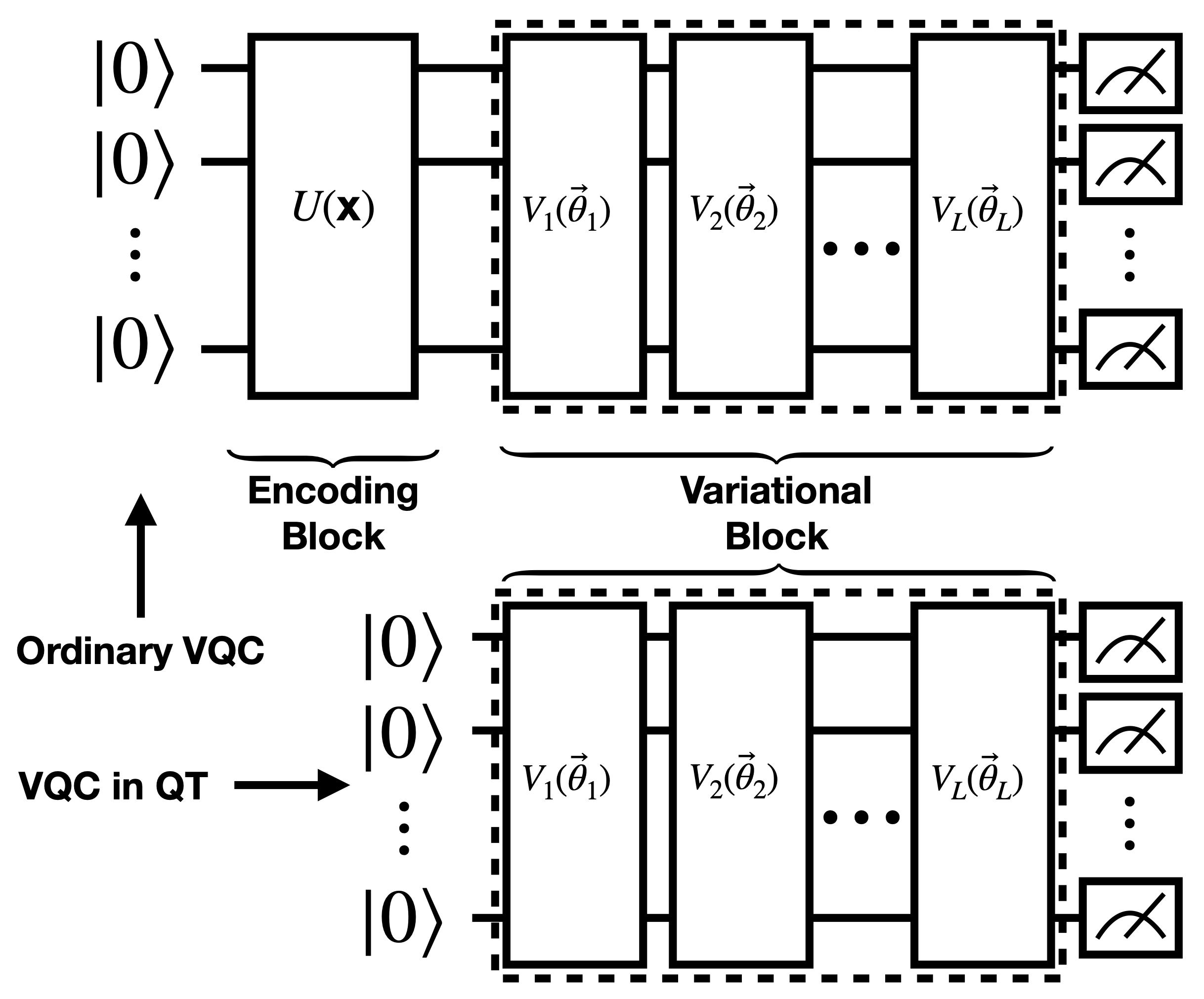}
\caption{{\bfseries Comparison between standard VQC and VQC used in the Quantum-Train (QT) framework.} The upper circuit illustrates a conventional variational quantum circuit, which consists of an encoding block $U(\vec{x})$ for input-dependent state preparation, followed by a variational block composed of layered parametrized unitaries $\{ V_\ell(\vec{\theta}_\ell)\}$. In contrast, the lower diagram shows the VQC design employed in QT, where the encoding circuit is omitted. By initializing directly from the ground state $\ket{0}^{\otimes n}$, the QT-based VQC reduces circuit depth and resource requirements, facilitating more efficient quantum execution.}
\label{fig:generic_VQC}
\end{center}
\end{figure}
The standard design of a VQC comprises three principal stages: (1) an input encoding circuit, (2) a sequence of parameterized quantum layers forming the variational core, and (3) a final measurement process. As depicted in the upper section of \figureautorefname{\ref{fig:generic_VQC}}, the encoding circuit $U(\mathbf{x})$ transforms a classical input vector $\mathbf{x}$ into a quantum state by acting on the initial all-zero qubit state $\ket{0}^{\otimes n}$, resulting in the embedded quantum state $\ket{\Psi} = U(\mathbf{x})\ket{0}^{\otimes n}$, where $n$ is the number of qubits involved.
Following the encoding stage, the quantum state is processed by a variational block-indicated by the dashed box-consisting of multiple layers $V_{j}(\vec{\theta_{j}})$ of trainable unitaries. Each layer may incorporate entangling operations (e.g., CNOT gates) as well as single-qubit parameterized rotations such as $R_{x}$, $R_{y}$ or $R_{z}$. The overall trainable module is denoted by $W(\Theta)$,  defined as the ordered product: 
\begin{equation}
    W(\Theta) = V_{L}(\Vec{\theta_{L}})V_{L-1}(\Vec{\theta_{L-1}}) \cdots V_{1}(\Vec{\theta_{1}}),
\end{equation}
where $L$ is the number of layers and $\Theta = \{\Vec{\theta_{1}}, \cdots, \Vec{\theta_{L}} \}$ collects all learnable parameters. Each vector $\Vec{\theta_{k}}$ contains rotation parameters for the $k$-th layer.
In the context of the \emph{Quantum-Train} (QT) framework, the VQC structure remains largely consistent with the conventional form, except for two key differences. First, the encoding block $U(\mathbf{x})$ is omitted, meaning the circuit operates directly from the ground state $\ket{0}^{\otimes n}$, thereby reducing circuit depth and resource overhead. Second, the measurement strategy diverges from the standard approach: while traditional VQCs typically compute expectation values of fixed observables such as Pauli-$Z$, yielding output vectors of the form
\begin{equation}
    \overrightarrow{f(\mathbf{x} ; \Theta)}=\left(\left\langle\hat{Z}_1\right\rangle, \cdots,\left\langle\hat{Z}_n\right\rangle\right),
\end{equation}
with each component given by 
\begin{equation}
    \left\langle\hat{Z}_{k}\right\rangle =\left\langle 0\left|U^{\dagger}(\mathbf{x})W^{\dagger}(\Theta) \hat{Z}_{k} W(\Theta)U(\mathbf{x})\right| 0\right\rangle,
\end{equation}
the QT module instead retrieves the full distribution over all $2^{n}$ computational basis states (e.g., $\ket{00 \cdots 0}, \cdots, \ket{11 \cdots 1}$). These measurement probabilities are subsequently utilized in downstream classical post-processing or model generation tasks.
\section{Quantum-Train}
\begin{figure}[htbp]
\vskip -0.1in
\begin{center}
\includegraphics[width=1\columnwidth]{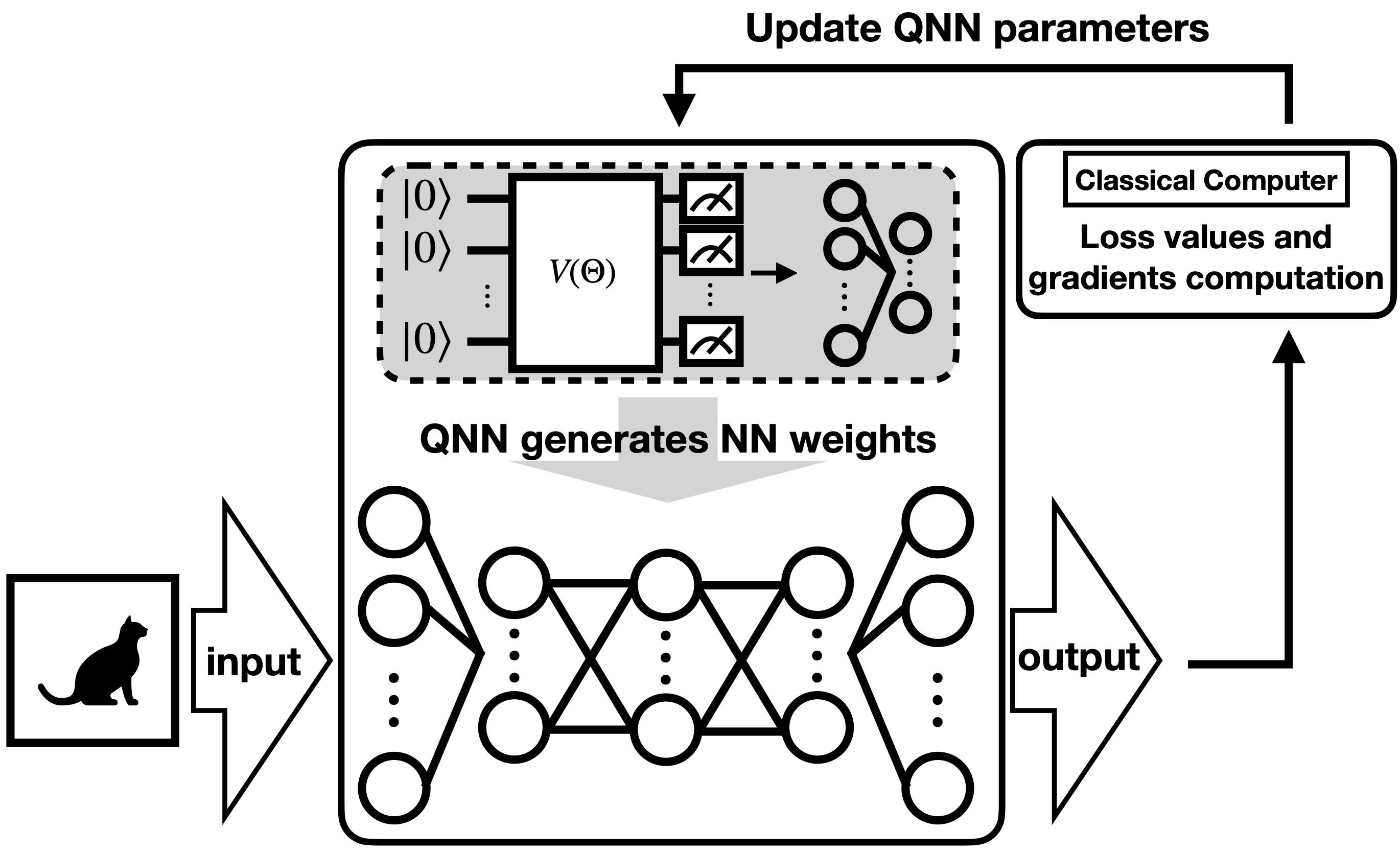}\vskip -0.1in
\caption{{\bfseries Concept of Quantum-Train.}}
\label{fig:QT_Concept}
\end{center}
\vskip -0.2in
\end{figure}

Given a target classical model such as a deep neural network with $p$ parameters, denoted $\kappa \in \mathbb{R}^{p}$, the Quantum-Train (QT) framework utilizes a quantum neural network (QNN) comprising $n_{qt} = \lceil \log_2 p \rceil$ qubits. This configuration defines a Hilbert space of dimension $2^{n_{qt}} \geq p$, enabling the generation of $2^{n_{qt}}$ measurement outcomes, each corresponding to a basis state $|\phi_i \rangle$, where $i \in {1, 2, \dots, 2^{n_{qt}}}$. The associated probabilities are given by $|\langle \phi_i | \psi(\gamma) \rangle|^2 \in [0,1]$, where $|\psi(\gamma)\rangle$ represents the QNN output state parameterized by $\gamma$. These probabilities are subsequently transformed into real-valued outputs via a classical mapping function $\mathcal{M}_{\beta}$, parameterized by $\beta$. Implemented as a multi-layer perceptron (MLP), $\mathcal{M}_{\beta}$ takes both the basis state label $|\phi_i\rangle$ (written in binary representation with length $n_{qt}$) and its corresponding measurement probability as input to generate the classical model parameter $\kappa_i$: 
\begin{equation} \mathcal{M}_{\beta}(|\phi_i \rangle, |\langle \phi_i | \psi (\gamma) \rangle|^2) = \kappa_i, \quad \forall i \in {1,2,\ldots,p}. \end{equation} 
Consequently, training the classical model parameters $\kappa$ entails joint optimization over the QNN parameters $\gamma$ and the mapping function parameters $\beta$. Provided that both $|\psi(\gamma)\rangle$ and $\mathcal{M}_\beta$ require only $O(\text{poly}(n_{qt}))$ parameters \cite{cerezo2021variational, sim2019expressibility}, this approach yields an effective compression from $p$ to $O(\text{polylog}(p))$ parameters. An overview of the QT mechanism is illustrated in \figureautorefname{\ref{fig:QT_Concept}}.
\section{Differentiable Quantum Architecture Search}
\label{sec:DiffQAS}
\begin{figure}[htbp]
\vskip -0.1in
\begin{center}
\includegraphics[width=1\columnwidth]{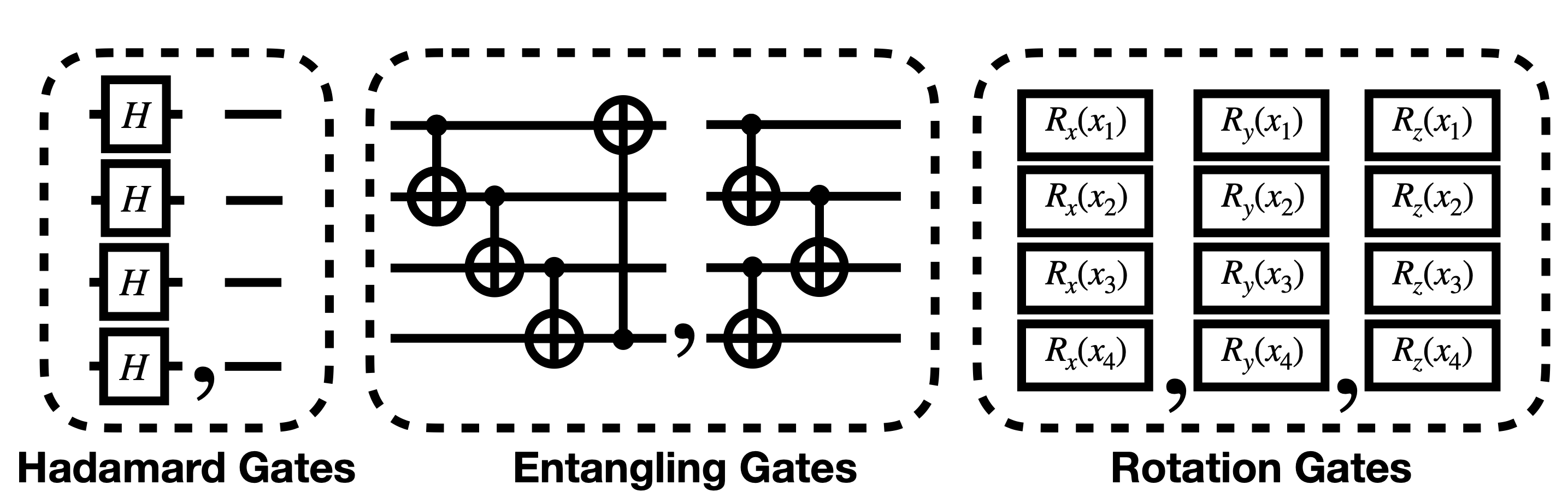}\vskip -0.1in
\caption{{\bfseries Ansatzes of QNN/VQC considered in this work.}}
\label{fig:Candidate_Circuits}
\end{center}
\vskip -0.2in
\end{figure}

\begin{figure}[htbp]
\vskip -0.1in
\begin{center}
\includegraphics[width=1\columnwidth]{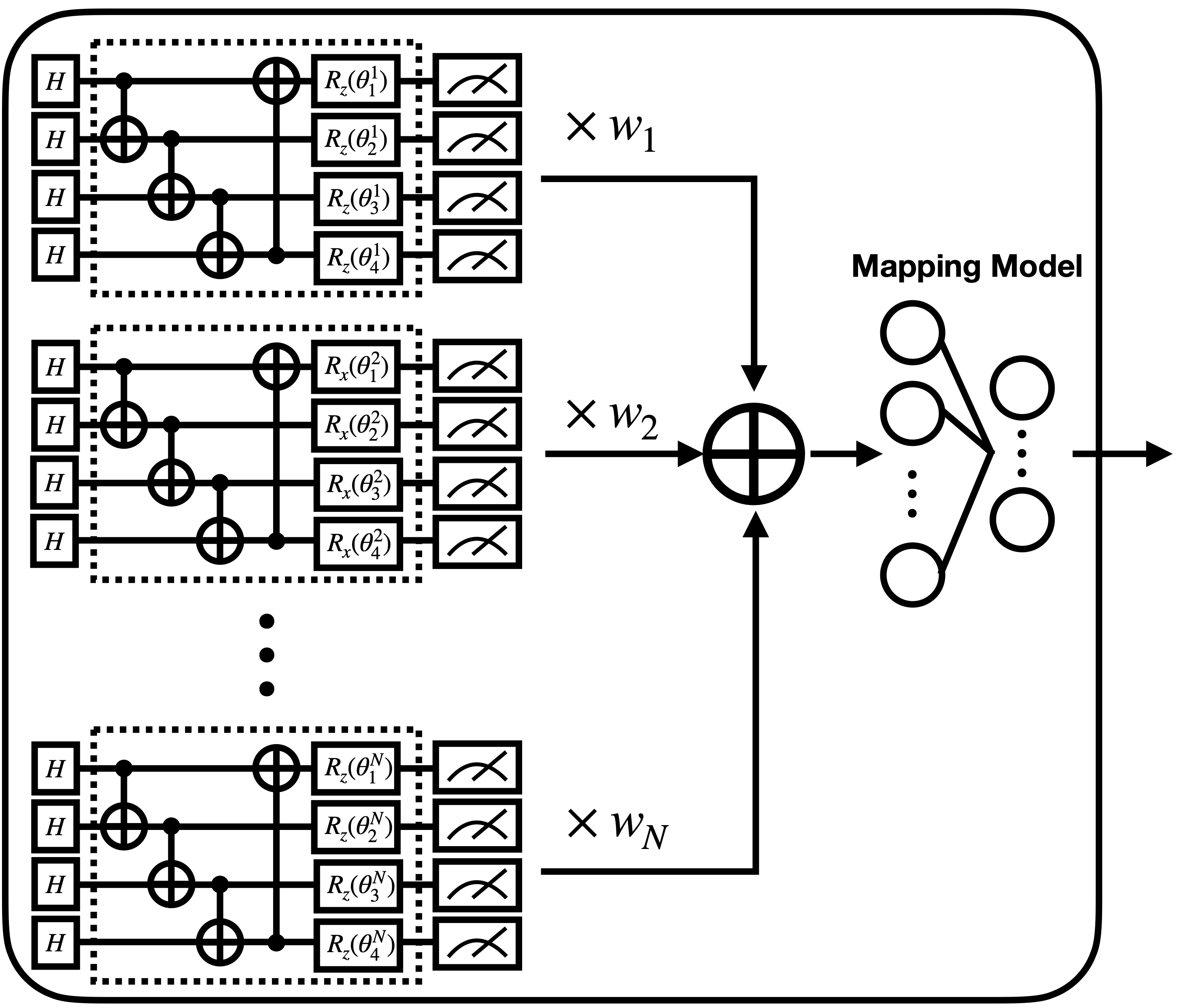}\vskip -0.1in
\caption{{\bfseries Quantum-Train with DiffQAS}}
\label{fig:DiffQAS_QT}
\end{center}
\vskip -0.2in
\end{figure}

Building upon ideas from classical neural architecture search (NAS) \cite{liu2018darts} and recent developments in differentiable quantum architecture search (DiffQAS) \cite{zhang2022differentiable,chen2024differentiable}, our method begins by defining a set of candidate quantum subcircuits. Suppose we aim to construct a quantum circuit $\mathcal{C}$, composed of multiple modular components $\mathcal{S}_{1}, \mathcal{S}_{2}, \dots, \mathcal{S}_{n}$. Each subcomponent $\mathcal{S}_{i}$ can be instantiated by selecting from a set of predefined circuit structures $\mathcal{B}_{i}$, and the cardinality $|\mathcal{B}_{i}|$ determines the number of possible designs at that position. Consequently, the total design space for $\mathcal{C}$ includes $N = |\mathcal{B}_{1}| \times |\mathcal{B}_{2}| \times \cdots \times |\mathcal{B}_{n}|$ unique circuit configurations.
For each candidate realization $\mathcal{C}_j$ within this space, we assign a learnable structural weight $w_j$, where $j \in {1, \dots, N}$. Each configuration $\mathcal{C}_{j}$ is also associated with a distinct set of trainable quantum parameters $\theta_j$. In a given machine learning task, some configurations may lead to well-performing models when trained, while others may prove ineffective. To integrate all possible candidates, we define an ensemble function $f_{\mathcal{C}}$ as a weighted combination over all $N$ candidates:
\begin{equation}
    f_{\mathcal{C}} = \sum_{j = 1}^{N} w_{j}f_{\mathcal{C}_{j}}
\end{equation}
where for brevity, we omit explicit references to the inputs $\vec{x}$ and internal parameters $\theta_j$.
The ensemble output $f_{\mathcal{C}}$ is then evaluated using a task-specific loss function $\mathcal{L}(f_{\mathcal{C}})$. By leveraging automatic differentiation, we compute gradients with respect to the structural weights $w_j$, i.e., $\nabla_{w_j} \mathcal{L}(f_{\mathcal{C}})$, allowing standard gradient-based optimization techniques to update the weights efficiently.
Within the DiffQAS framework, variational quantum circuits (VQCs) are constructed by selecting from a set of pre-defined ansatz candidates, as depicted in \figureautorefname{\ref{fig:Candidate_Circuits}}. Unlike traditional QNNs, the Quantum-Train (QT) setting omits an explicit encoding circuit. Instead, the architecture is formed by choosing among three design dimensions: (1) whether to apply Hadamard gates at initialization to generate unbiased superpositions, (2) which entangling layer to use (two choices are provided), and (3) the type of rotation gates to be applied ($R_x$, $R_y$, or $R_z$). This results in $2 \times 2 \times 3 = 12$ total architecture candidates for a single-layer VQC or QNN, each assigned an individual weight $w_j$.
To enhance expressiveness and accommodate more complex tasks, multiple ensemble layers can be stacked to build deeper quantum circuits. If $M$ such layers are employed, the total number of structural weights increases to $N \times M$, reflecting the full space of configurable subcircuits across all layers. All structural weights across the depth of the circuit can be jointly optimized through differentiable programming techniques and standard optimizers.
The overall concept of integrating Quantum-Train with DiffQAS is illustrated in \figureautorefname{\ref{fig:DiffQAS_QT}}. Based on the set of allowed circuit components defined in \figureautorefname{\ref{fig:Candidate_Circuits}}, we construct all 12 feasible circuit configurations. During each forward pass, all candidate circuits—implemented as $n$-qubit QNNs—independently execute their quantum computations, each producing $2^n$ measurement probabilities corresponding to the computational basis states.
These outputs are subsequently aggregated via a weighted summation governed by the trainable structural weights $w_j$, yielding the final ensemble output:
\begin{equation}
    \text{output} = \sum w_{j} \text{VQC}_{j}(\Theta^{j})
\end{equation}
where $\text{VQC}_j(\Theta^{j})$ denotes the output from the $j$-th variational quantum circuit with its own parameter set $\Theta^{j}$.
While assigning distinct parameter sets $\Theta^{j}$ to each candidate architecture is theoretically permissible, doing so significantly increases the total number of trainable parameters. To mitigate this complexity, we adopt a shared parameter strategy in this study: a single parameter vector $\Theta$ is used across all circuit variants. Accordingly, the ensemble output from the DiffQAS block simplifies to:
\begin{equation}
    \text{output} = \sum w_{j} \text{VQC}_{j}(\Theta).
\end{equation}
\section{Numerical Results and Discussion}
\label{sec:nrd}
To assess the effectiveness of the proposed DiffQAS-QT framework, we conduct comprehensive evaluations across three categories of tasks: classification, time-series prediction, and reinforcement learning. Specifically, we benchmark performance on the MNIST and FashionMNIST datasets for classification; the Bessel function, damped simple harmonic motion (SHM), delayed quantum control, and NARMA-5/10 tasks for time-series prediction; and the MiniGrid-Empty environments (5×5 and 6×6) for reinforcement learning. In our comparisons, we include 12 baseline quantum neural network (QNN) configurations, detailed in \tableautorefname{\ref{tab:baseline_configs}}. For the entanglement strategy within the circuit design, we consider two architectural variants: the \emph{Ent} design corresponds to the right-hand configuration illustrated in \figureautorefname{\ref{fig:Candidate_Circuits}}, whereas the \emph{Cyc} design follows the left-hand configuration in the same figure.
\begin{table}[htbp]
\caption{QNN Baseline Configs}
\label{tab:baseline_configs}
\resizebox{\columnwidth}{!}{%
\begin{tabular}{|l|l|l|l|l|l|l|l|l|l|l|l|l|}
\hline
\diaghead{\theadfont Diag ColumnmnHead II}%
  {Component}{VQC config} & 1       & 2       & 3       & 4       & 5       & 6       & 7       & 8       & 9       & 10      & 11      & 12      \\ \hline
Hadamard Gate                                                                                          & Y       & Y       & Y       & Y       & Y       & Y       & N       & N       & N       & N       & N       & N       \\ \hline
Entanglement                                                                                           & Ent     & Ent     & Ent     & Cyc     & Cyc     & Cyc     & Ent     & Ent     & Ent     & Cyc     & Cyc     & Cyc     \\ \hline
Rotation                                                                                               & $R_{x}$ & $R_{y}$ & $R_{z}$ & $R_{x}$ & $R_{y}$ & $R_{z}$ & $R_{x}$ & $R_{y}$ & $R_{z}$ & $R_{x}$ & $R_{y}$ & $R_{z}$ \\ \hline
\end{tabular}%
}
\end{table}
\subsection{Classification}
\subsubsection{Simulation Setting}
For the classification tasks, we consider three binary settings: MNIST (digits 1 vs 5), FashionMNIST (classes 1 vs 5), and FashionMNIST (classes 5 vs 7). All models are trained using a batch size of 100 and the Adam optimizer with a learning rate of $10^{-3}$. The QNN depth is set to 15 layers, as shallower circuits (i.e., with fewer than 15 layers) were found to be insufficient for successful training in preliminary experiments. In the conventional neural network setting, the total number of trainable parameters reaches 108{,}866. By contrast, the proposed DiffQAS-QT framework reduces this dramatically to only 286 trainable parameters, demonstrating the potential for substantial model compression while maintaining learnability.
\subsubsection{Result Analysis}
\begin{figure}[htbp]
\begin{center}
\includegraphics[width=1\columnwidth]{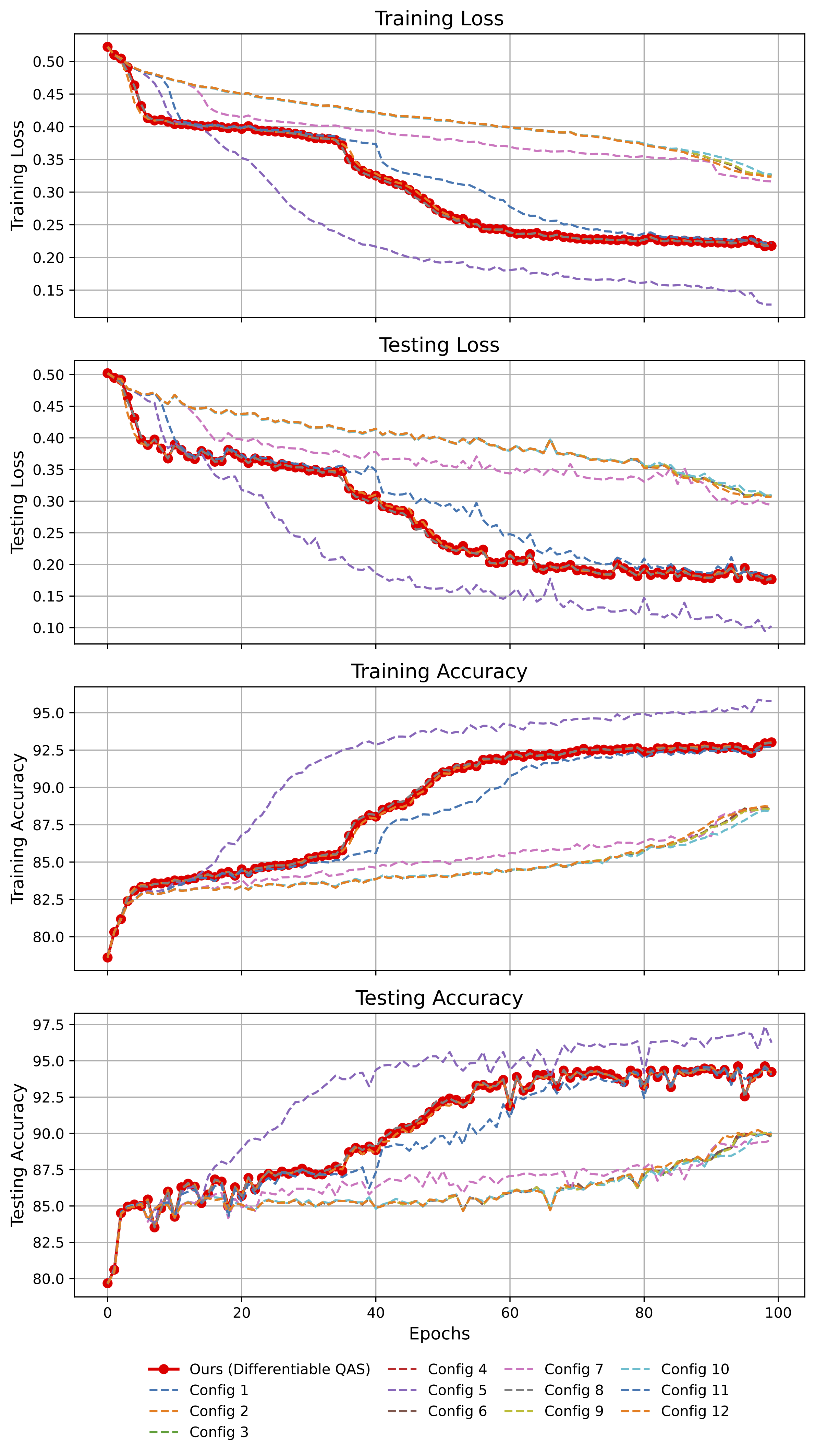}
\caption{{\bfseries Comparison of Various QNN Architectures on Binary Classification of Digits 1 and 5 from the MNIST Dataset.}}
\label{fig:results_classification_MNIST_1_vs_5}
\end{center}
\end{figure}

\begin{figure}[htbp]
\begin{center}
\includegraphics[width=1\columnwidth]{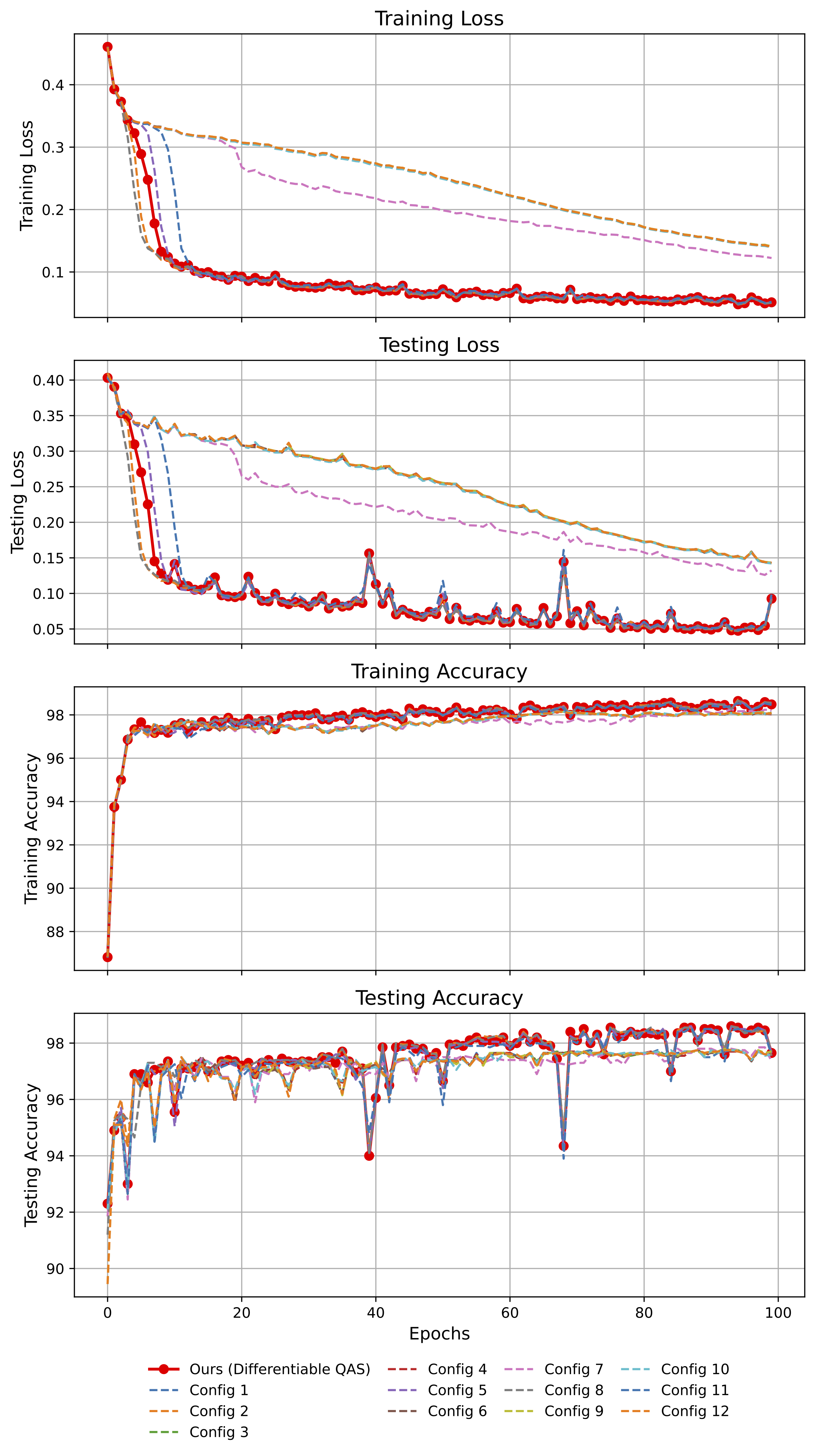}
\caption{{\bfseries Comparison of QNN Architectures on FashionMNIST Binary Classification (Labels 1 vs 5).}}
\label{fig:results_classification_FashionMNIST_1_vs_5}
\end{center}
\end{figure}

\begin{figure}[htbp]
\begin{center}
\includegraphics[width=1\columnwidth]{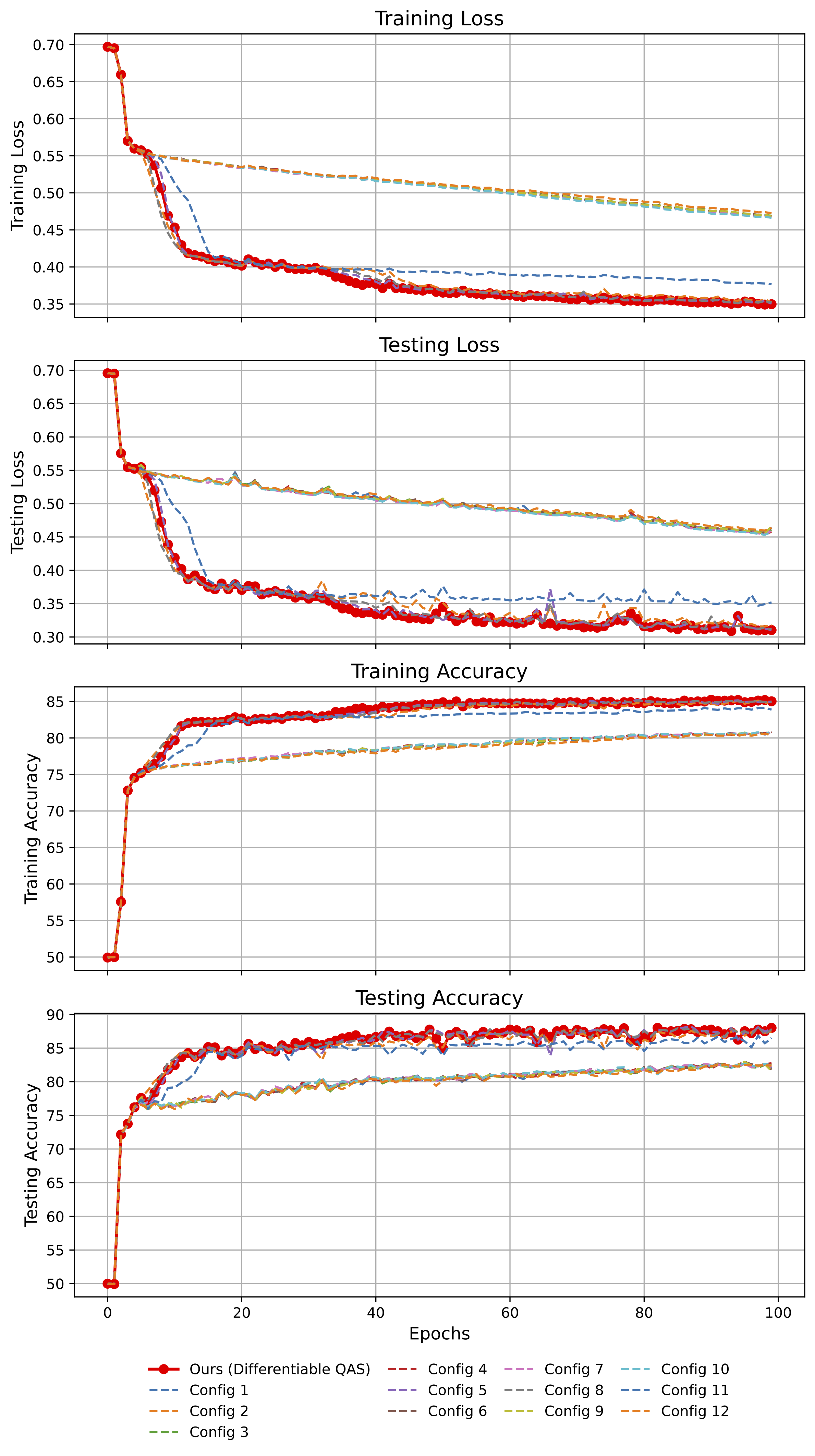}
\caption{{\bfseries Comparison of QNN Architectures on FashionMNIST Binary Classification (Labels 5 vs 7).}}
\label{fig:results_classification_FashionMNIST_5_vs_7}
\end{center}
\end{figure}

We conduct an initial evaluation of the proposed DiffQAS-QT approach on the MNIST dataset for binary classification between digits 1 and 5. As shown in \figureautorefname{\ref{fig:results_classification_MNIST_1_vs_5}}, DiffQAS-QT achieves performance close to the best manually-crafted QNN architectures, with slightly lower peak testing accuracy but significantly improved stability and generalization. The model exhibits smooth convergence with steadily decreasing loss and increasing accuracy, highlighting its robustness and efficiency despite being automatically discovered via differentiable architecture search.
We further evaluate the proposed DiffQAS-QT framework on the FashionMNIST dataset for binary classification of classes 1 and 5. As shown in \figureautorefname{\ref{fig:results_classification_FashionMNIST_1_vs_5}}, the DiffQAS-QT model achieves consistently strong performance, with training and testing accuracy exceeding 98\% and comparable to the best manually-designed baselines. Although certain baseline configurations achieve faster initial convergence, the DiffQAS-QT model-automatically discovered via differentiable architecture optimization-demonstrates notable stability and low variance throughout training, especially in terms of testing accuracy. The model also maintains low training and testing loss throughout, indicating effective learning and strong generalization in more challenging visual domains.
Finally, we assess the performance of DiffQAS-QT on the FashionMNIST dataset for binary classification between labels 5 and 7, which are known to be visually similar and more challenging to separate. As illustrated in \figureautorefname{\ref{fig:results_classification_FashionMNIST_5_vs_7}}, the proposed model consistently outperforms most baseline configurations in both training and testing accuracy. While several baselines exhibit early rapid gains, DiffQAS-QT achieves higher final accuracy with lower testing loss and smoother convergence. These results highlight the ability of the automatically discovered architecture to generalize effectively even in fine-grained classification scenarios.
\subsection{Time-Series Prediction}
\subsubsection{Simulation Setting}
For the time-series prediction tasks, we evaluate the DiffQAS-QT framework on five representative problems: the Bessel function $J_{2}$, damped simple harmonic motion (SHM), delayed quantum control, and nonlinear autoregressive moving average models NARMA-5 and NARMA-10. In this setup, the DiffQAS-QT module is used to generate the parameters of an LSTM network. Training is conducted using a batch size of 10 and a QNN depth of 10. We adopt the RMSProp optimizer with a learning rate of 0.01, decay factor (alpha) of 0.99, and epsilon set to $10^{-8}$. The conventional LSTM baseline contains 1,781 trainable parameters, while the DiffQAS-QT-based version reduces this to just 135, illustrating the framework’s ability to achieve efficient parameterization with minimal resource overhead. To prepare the dataset, we employ a sliding window strategy. Each input sequence is composed of four consecutive values $[x_{t-4}, x_{t-3}, x_{t-2}, x_{t-1}]$, and the model is trained to predict the next value $x_{t}$. This window length is consistent with prior settings used in our QLSTM experiments \cite{chen2022quantumLSTM}, providing a coherent basis for comparison.
\subsubsection{Result Analysis}
\begin{figure}[htbp]
\begin{center}
\includegraphics[width=1\columnwidth]{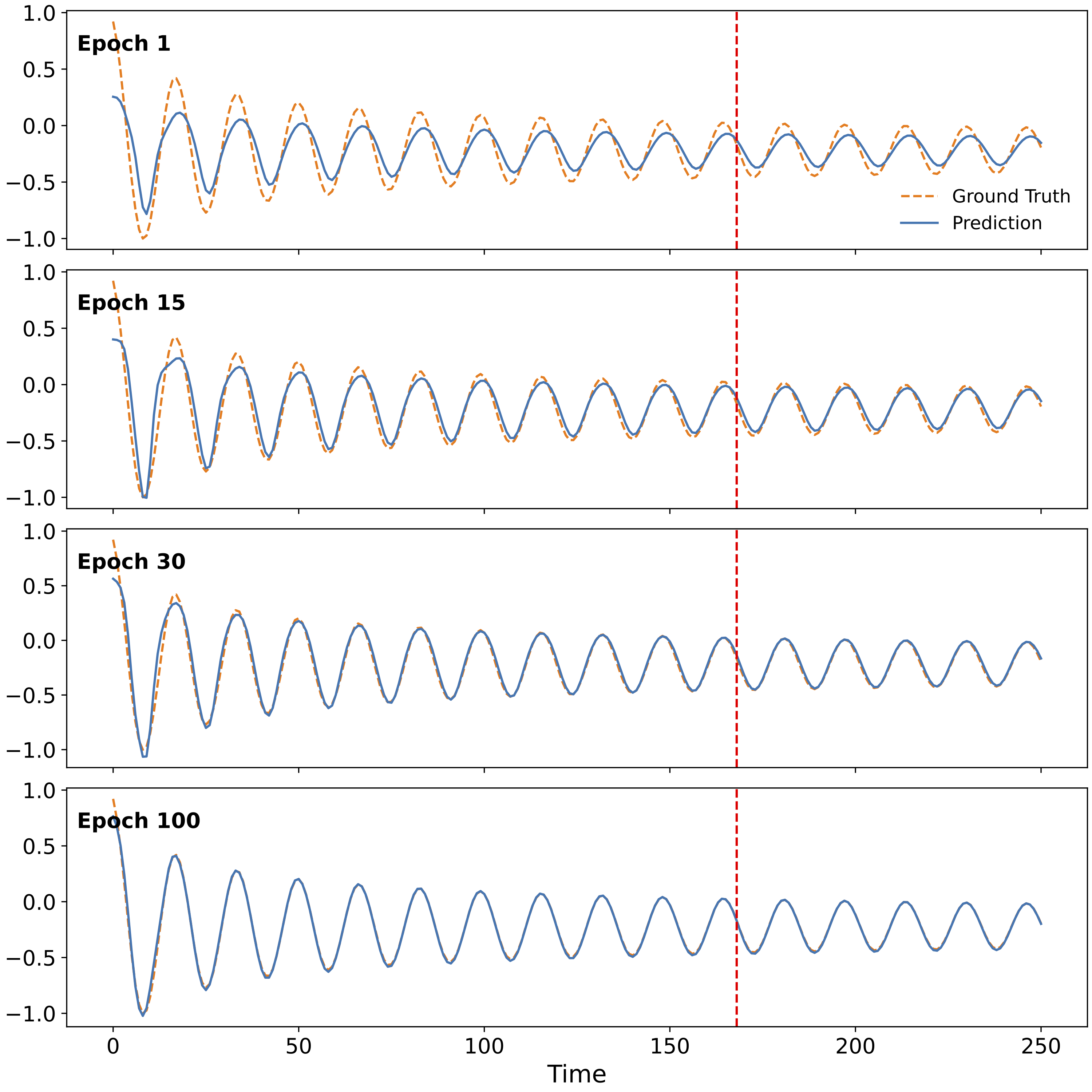}
\caption{{\bfseries Model Rollout on Bessel Function at Different Training Epochs.} The red dashed line separates the training region (left) from the testing region (right). The model's prediction (blue) improves over time in approximating the ground truth (orange), with increasing alignment beyond the training horizon from epoch 1 to epoch 100.}
\label{fig:results_Bessel_Rollout}
\end{center}
\end{figure}

\begin{figure}[htbp]
\begin{center}
\includegraphics[width=1\columnwidth]{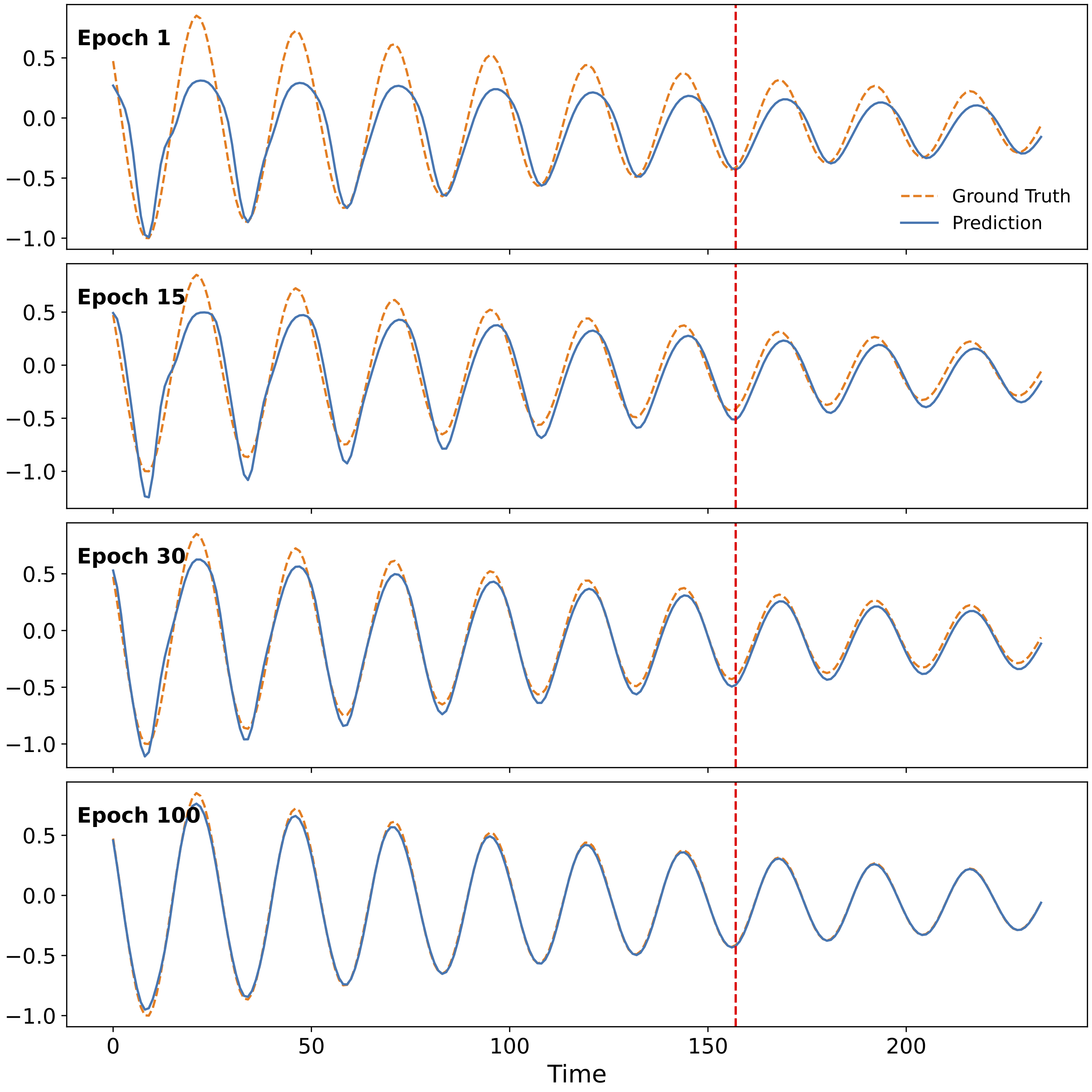}
\caption{{\bfseries Model Rollout on Damped SHM at Different Training Epochs.} Same model architectures and training epochs as in the previous figure, trained and evaluated on a different time series.}
\label{fig:results_Damped_SHM_Rollout}
\end{center}
\end{figure}

\begin{figure}[htbp]
\begin{center}
\includegraphics[width=1\columnwidth]{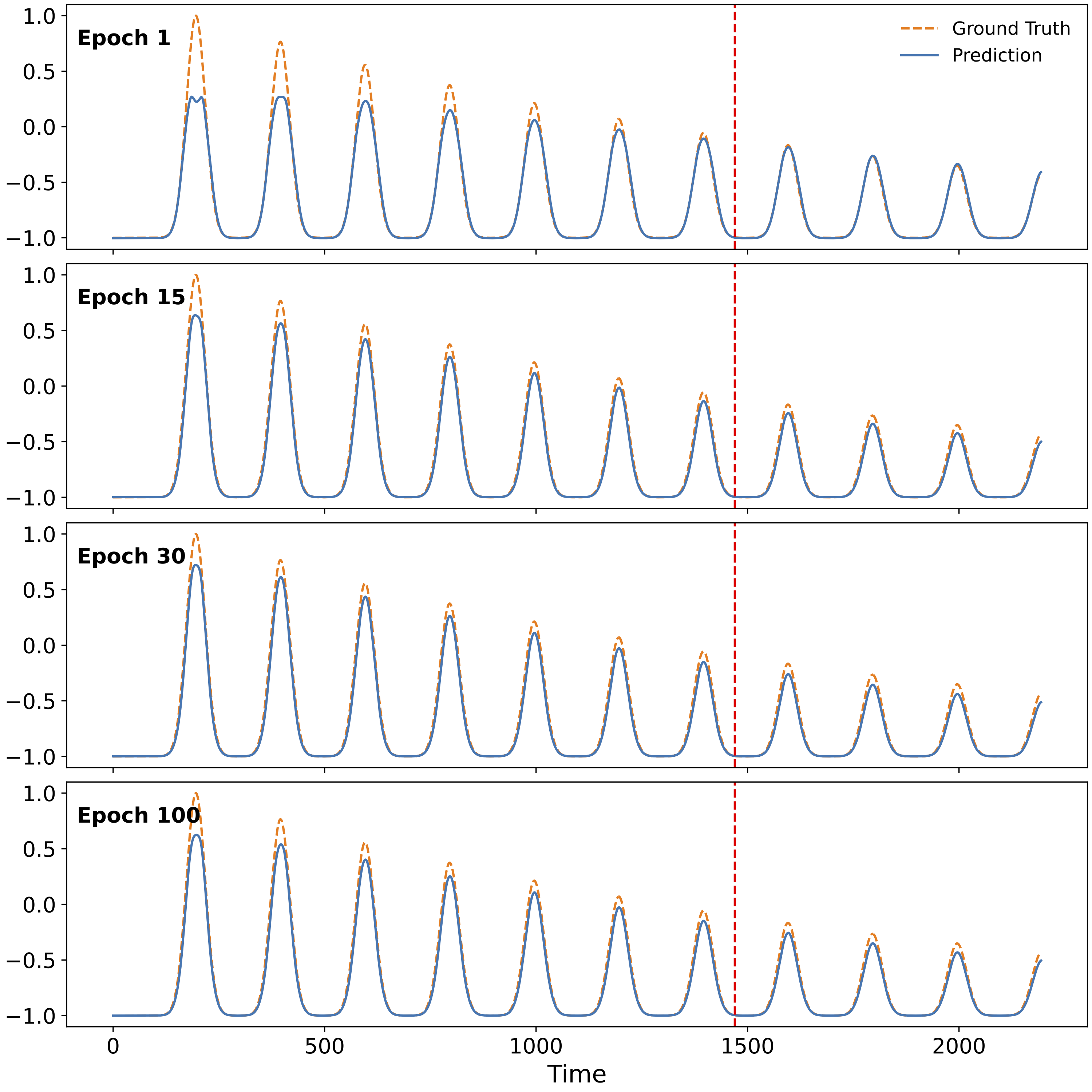}
\caption{{\bfseries Model Rollout on Delayed Quantum Control at Different Training Epochs.} Same configuration, different time series instance.}
\label{fig:results_Delayed_Quantum_Control_Rollout}
\end{center}
\end{figure}

\begin{figure}[htbp]
\begin{center}
\includegraphics[width=1\columnwidth]{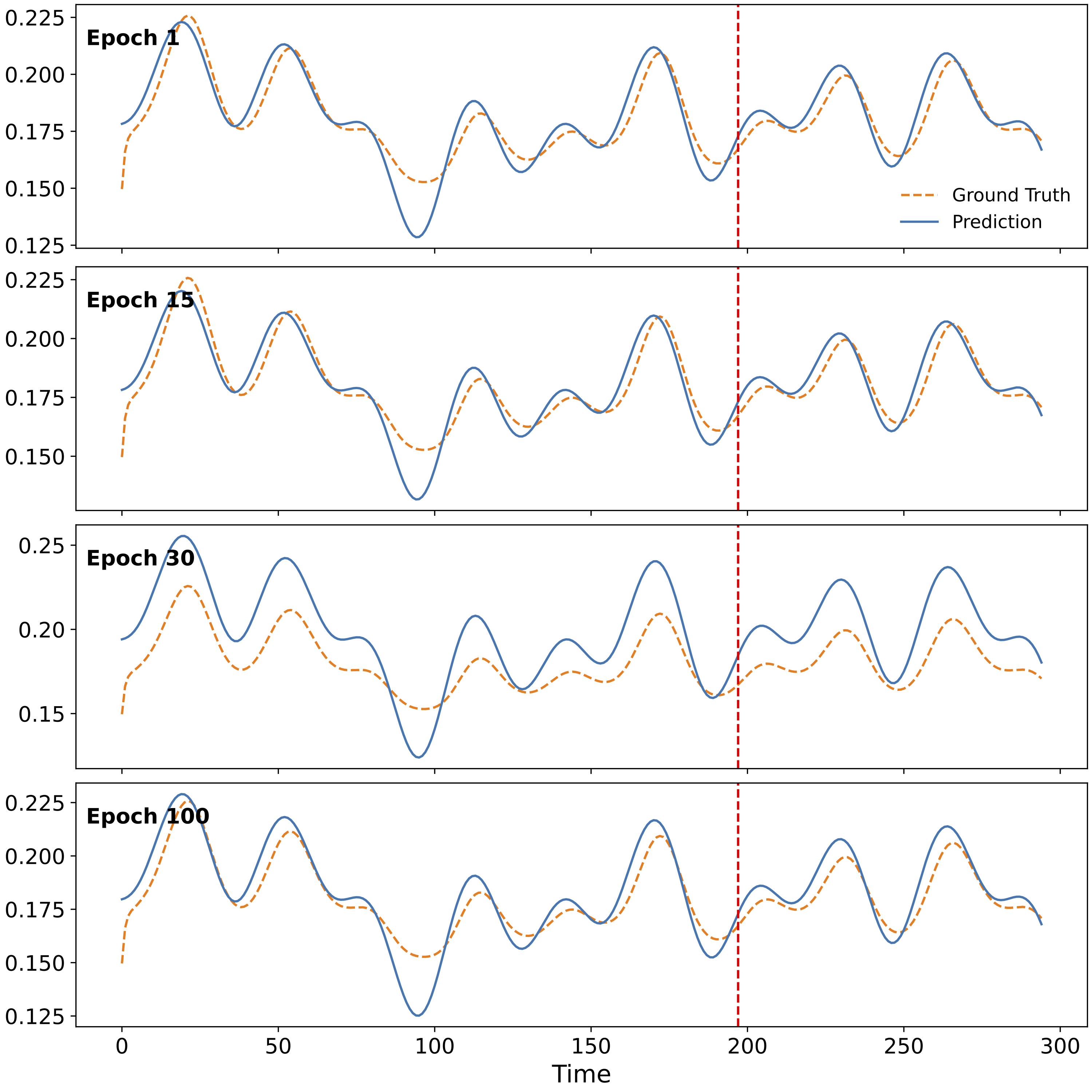}
\caption{{\bfseries Model Rollout on NARMA 5 at Different Training Epochs.} Another test sequence; consistent improvement in testing region.}
\label{fig:results_NARMA_5_Rollout}
\end{center}
\end{figure}

\begin{figure}[htbp]
\begin{center}
\includegraphics[width=1\columnwidth]{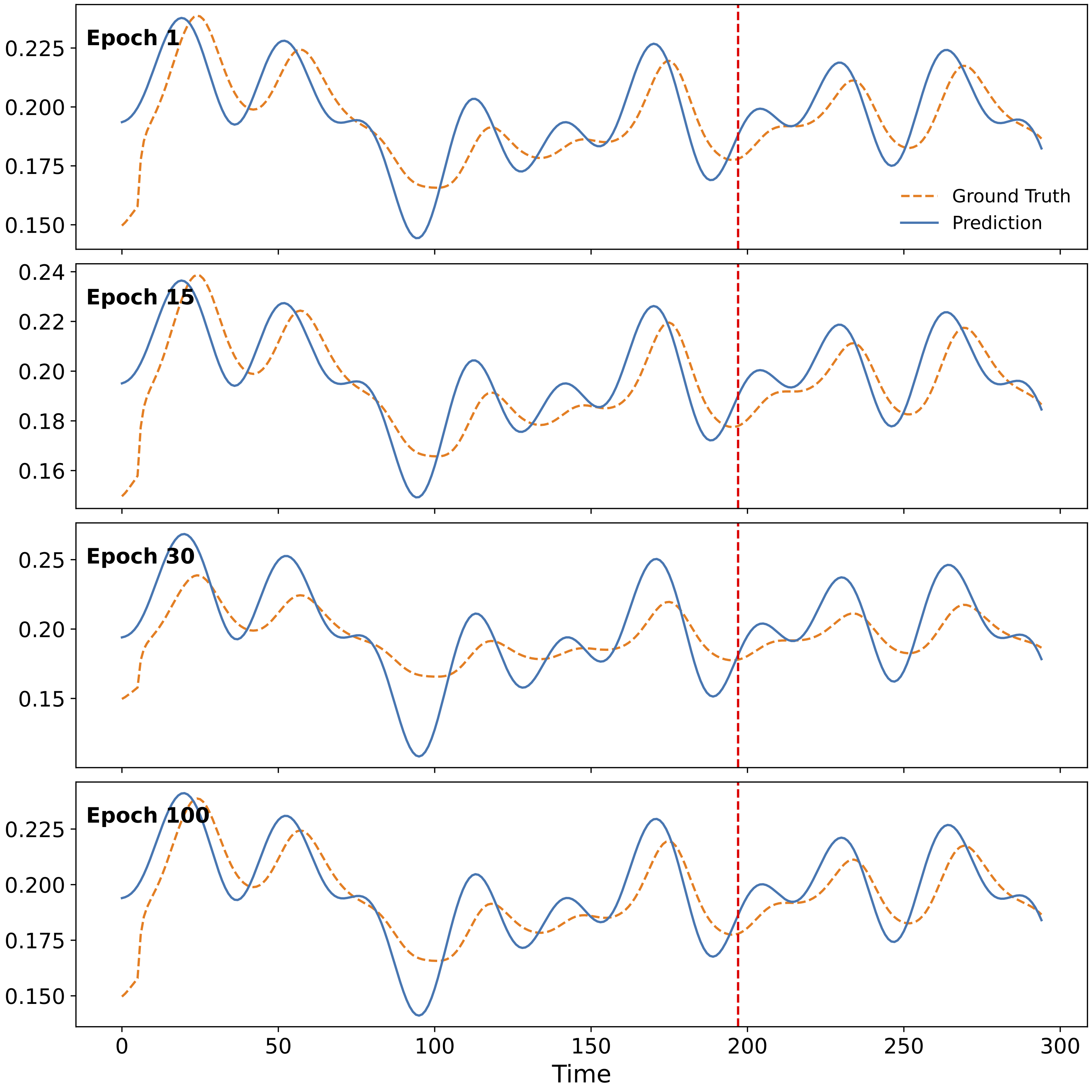}
\caption{{\bfseries Model Rollout on NARMA 10 at Different Training Epochs.} Model exhibits stable learning and prediction across varied sequences.}
\label{fig:results_NARMA_10_Rollout}
\end{center}
\end{figure}

We further evaluate the learned model in a time-series prediction task using the Bessel function ($J_{2}$) as the target signal. As shown in \figureautorefname{\ref{fig:results_Bessel_Rollout}}, we visualize the model’s predictions at four representative training epochs: 1, 15, 30, and 100. The red dashed line marks the boundary between the training region (left) and the testing region (right). As training progresses, the model exhibits increasingly accurate prediction behavior, with tighter alignment to the ground truth (orange dashed line) beyond the training horizon. By epoch 100, the prediction closely follows the ground truth across both the training and testing regions, indicating strong learning of the underlying temporal dynamics.
We further evaluate the model on a time-series prediction task using a damped simple harmonic motion (Damped SHM) signal. As shown in \figureautorefname{\ref{fig:results_Damped_SHM_Rollout}}, the model is trained directly on this sequence and evaluated at four representative training epochs: 1, 15, 30, and 100. The red dashed line separates the training region from the testing region. Over time, the model learns to accurately capture the decaying oscillatory pattern, with prediction trajectories (blue) aligning closely with the ground truth (orange dashed line) by epoch 100. These results indicate the model's ability to effectively learn and predict time-series signals with smooth, damped periodic structures.
We further evaluate the model on a time-series prediction task inspired by delayed quantum control. The underlying physical system consists of a two-level atom (qubit) coupled to a semi-infinite one-dimensional waveguide terminated by a perfect mirror. The mirror induces complete reflection of propagating photons, creating delayed quantum feedback. When the qubit–mirror separation $L$ is an integer multiple of the qubit’s resonant wavelength $\lambda_0$, a bound state in the continuum (BIC) forms, resulting in persistent photon trapping between the qubit and the mirror \cite{calajo2019exciting,tufarelli2013dynamics,dong2009intrinsic}. In this setting, we consider a sinusoidal modulation of the qubit frequency such that its time-averaged value satisfies the resonance condition. This periodic detuning causes the intermittent release of the trapped photon population, producing a measurable, temporally structured emission pattern \cite{tufarelli2013dynamics}. As shown in \figureautorefname{\ref{fig:results_Delayed_Quantum_Control_Rollout}}, the model is trained to learn this non-Markovian quantum behavior \cite{tufarelli2013dynamics,tufarelli2014non,fang2018non} directly from the output field intensity. Across training epochs 1, 15, 30, and 100, the predicted output (blue) aligns increasingly well with the ground truth signal (orange dashed), including accurate reproduction of both emission timing and amplitude. By epoch 100, the model exhibits high-fidelity prediction over the testing region, indicating successful learning of the underlying delayed feedback dynamics.
We further evaluate the model’s ability to capture nonlinear temporal dependencies using the NARMA benchmark tasks. Specifically, we consider both the NARMA 5 and NARMA 10 systems, which introduce increasing levels of long-range memory and nonlinear interactions between input and output sequences. As shown in \figureautorefname{\ref{fig:results_NARMA_5_Rollout}} and \figureautorefname{\ref{fig:results_NARMA_10_Rollout}}, the model is trained on each system independently and evaluated at multiple training epochs. In both cases, the predicted output (blue) exhibits consistent improvement in the testing region over time, gradually aligning with the ground truth (orange dashed). While NARMA 10 poses a greater challenge due to its longer memory horizon and more complex dynamics, the model demonstrates stable prediction behavior and improved fidelity by epoch 100. These results highlight the model’s ability to handle nonlinear dynamics with varying degrees of temporal complexity.
To quantitatively assess the model’s performance across diverse temporal tasks, we report the mean squared error (MSE) on five representative time-series datasets in \tableautorefname{\ref{tab:multi_task_mse}}. The proposed DiffQAS-based QT-LSTM model consistently achieves competitive or superior performance across all benchmarks. It obtains the lowest MSE on the Bessel and damped SHM datasets, and remains within the top-performing group on delayed quantum control and both NARMA tasks. Notably, while some baseline configurations attain slightly lower error on specific tasks (e.g., Config 5 on NARMA 5, Config 8 on NARMA 10), they often fail to generalize across other datasets. In contrast, the proposed method demonstrates stable and robust performance under varied temporal dynamics, ranging from smooth decay and oscillations to non-Markovian quantum feedback and high-order autoregressive dependencies.
\begin{table}[htbp]
\centering
\caption{Multi-task MSE Comparison Across Time-Series Datasets}
\label{tab:multi_task_mse}
\resizebox{\columnwidth}{!}{%
\begin{tabular}{lccccc}
\toprule
\textbf{Model} & \textbf{Bessel} $\downarrow$ & \textbf{Damped SHM} $\downarrow$ & \textbf{Delayed QC} $\downarrow$ & \textbf{NARMA 5} $\downarrow$ & \textbf{NARMA 10} $\downarrow$ \\
\midrule
Ours & \textbf{0.000040} & \textbf{0.000036} & 0.001710 & 0.000050 & 0.000117 \\
Config 1 & 0.000515 & 0.000155 & 0.000715 & 0.000038 & 0.000095 \\
Config 2 & 0.000525 & 0.000695 & 0.001190 & 0.000068 & 0.000103 \\
Config 3 & 0.000515 & 0.000155 & 0.000676 & 0.000038 & 0.000095 \\
Config 4 & 0.000515 & 0.000155 & 0.002011 & 0.000038 & 0.000095 \\
Config 5 & 0.002902 & 0.000820 & 0.001111 & \textbf{0.000024} & 0.000098 \\
Config 6 & 0.000515 & 0.000155 & 0.002638 & 0.000038 & 0.000095 \\
Config 7 & 0.000601 & 0.000040 & 0.001679 & 0.000052 & 0.000120 \\
Config 8 & 0.000141 & 0.008873 & 0.001723 & 0.000091 & \textbf{0.000081} \\
Config 9 & 0.001391 & 0.000154 & 0.000956 & 0.000045 & 0.000125 \\
Config 10 & 0.000216 & 0.000157 & 0.001839 & 0.000040 & 0.000121 \\
Config 11 & 0.007428 & 0.032072 & \textbf{0.000211} & 0.000054 & 0.000140 \\
Config 12 & 0.001391 & 0.000154 & 0.000956 & 0.000045 & 0.000125 \\
\bottomrule
\end{tabular}%
}
\end{table}

\subsection{Reinforcement Learning}
\subsubsection{Simulation Setting}
We further evaluate the DiffQAS-QT framework in the reinforcement learning (RL) setting using the MiniGrid-Empty environments of size 5×5 and 6×6 \cite{gym_minigrid}. Training is conducted using the Asynchronous Advantage Actor-Critic (A3C) algorithm \cite{mnih2016asynchronous, chen2023asynchronous}, with 16 parallel worker processes. The discount factor is set to $\gamma = 0.9$, and and model updates occur every 5 steps. For the quantum modules, we set the QNN depth to 10 and use the Adam optimizer with a learning rate of $10^{-4}$ and betas $(0.92, 0.999)$. In the standard baseline configuration, the policy network and value network contain 6,023 and 5,825 trainable parameters, respectively. In contrast, under the DiffQAS-QT framework, each network incorporates its own DiffQAS-QT module comprising only 157 trainable parameters. This results in a combined total of just 314 parameters across both networks, highlighting a substantial reduction in model complexity.
\subsubsection{Result Analysis}
\begin{figure}[htbp]
\begin{center}
\includegraphics[width=1\columnwidth]{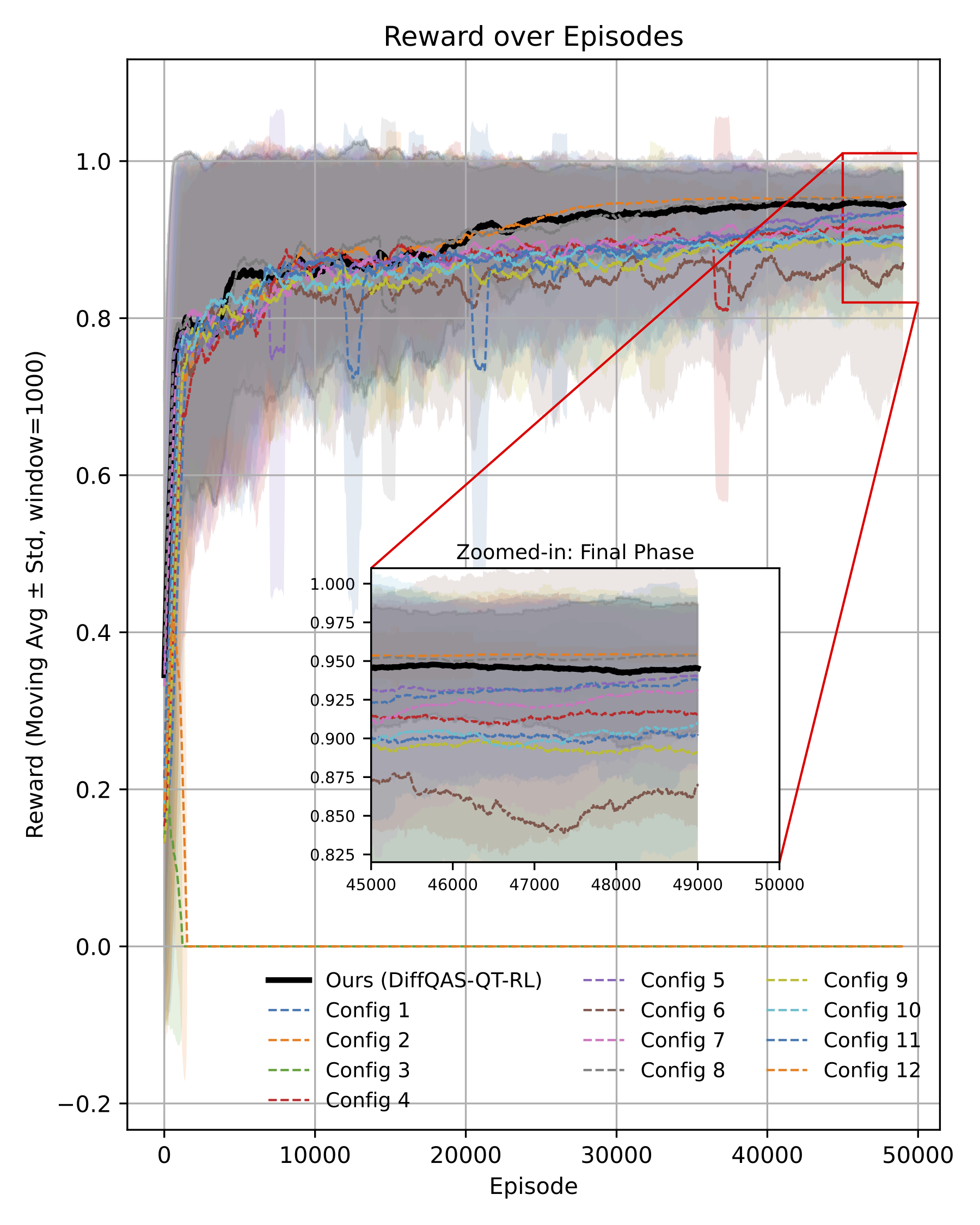}
\caption{{\bfseries Reward Curve Comparison on MiniGrid-Empty-5x5 Environment.} Episodic rewards (moving average ± standard deviation, window = 1000) over 50,000 episodes are shown for our method (DiffQAS-QT-RL, black) and 12 manually-designed baseline configurations. The red inset zooms into the final 5,000 episodes, where our method consistently demonstrates higher rewards and greater stability compared to most baselines.}
\label{fig:results_MiniGrid_Empty_5x5}
\end{center}
\end{figure}

\begin{figure}[htbp]
\begin{center}
\includegraphics[width=1\columnwidth]{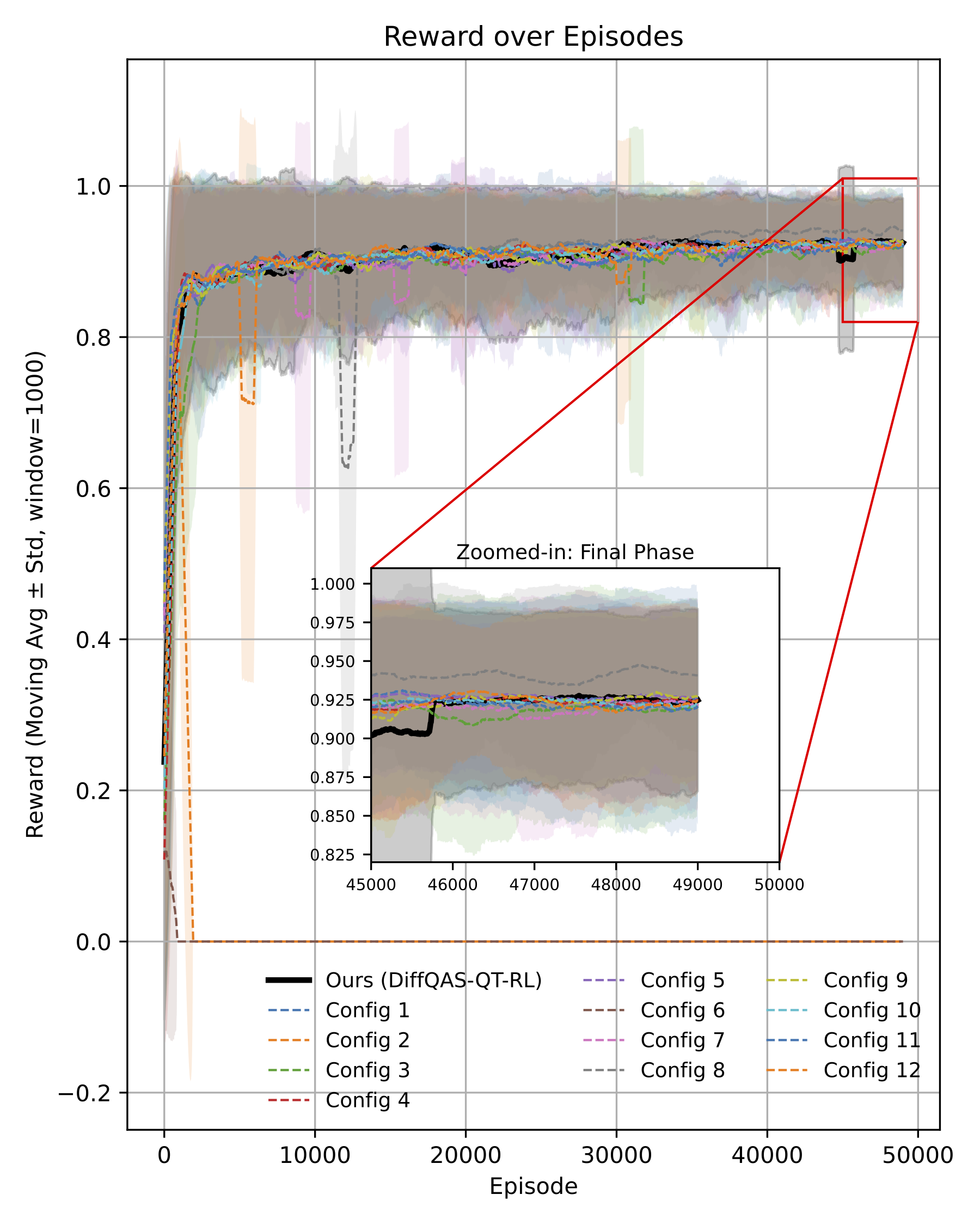}
\caption{{\bfseries Reward Curve Comparison on MiniGrid-Empty-6x6 Environment.} Results on the larger 6x6 environment show trends consistent with the 5x5 case. Our method (DiffQAS-QT-RL) maintains superior reward and stability over most handcrafted baselines, indicating robustness to increased spatial complexity.}
\label{fig:results_MiniGrid_Empty_6x6}
\end{center}
\end{figure}

We evaluate the proposed DiffQAS-QT-RL model in a reinforcement learning setting using the MiniGrid-Empty-5x5 environment. \figureautorefname{\ref{fig:results_MiniGrid_Empty_5x5}} shows the episodic rewards (moving average ± standard deviation, window = 1000) across 50,000 episodes, compared to 12 manually-designed QNN baselines. The proposed model (black curve) achieves higher average rewards and significantly lower variance during the final training phase. The inset highlights the last 5,000 episodes, where our method consistently outperforms most baselines in both performance and stability. These results suggest that the differentiable quantum architecture search framework can effectively discover robust quantum-enhanced policies for navigation tasks in grid-based environments.
To assess the robustness of our method under increased spatial complexity, we further evaluate it on the MiniGrid-Empty-6x6 environment. As shown in \figureautorefname{\ref{fig:results_MiniGrid_Empty_6x6}}, the reward learning curves maintain trends consistent with the 5x5 case. The DiffQAS-QT-RL model continues to outperform most handcrafted baselines in both final reward and stability, especially during the late-stage episodes as highlighted in the zoomed-in inset. These results suggest that our architecture search framework is capable of discovering quantum-enhanced policies that scale well with task complexity, retaining strong performance under more demanding exploration and navigation conditions.
\section{Conclusion and Future Work}
\label{sec:cfw}
This paper demonstrates the effectiveness of applying differentiable quantum architecture search (DiffQAS) for constructing quantum-enhanced neural network programmer, referred to as quantum-train (QT), across three fundamental learning paradigms—supervised classification, time-series prediction, and reinforcement learning. Our results show that QNN architectures discovered via DiffQAS not only achieve performance competitive with or superior to handcrafted baselines, but also generalize well across tasks with varying temporal and spatial complexity.
The broad applicability of DiffQAS-QT across supervised, temporal, and interactive settings suggests that differentiable quantum architecture search may serve as a core engine for developing versatile, high-performing quantum AI systems.

\section*{Acknowledgment}
Wei-Hao Huang (Jij inc) acknowledges this work was performed for the Council for Science, Technology and
Innovation (CSTI), Cross-ministerial Strategic Innovation Promotion Program (SIP), ``Promoting the application of advanced quantum technology platforms to social issues'' (Funding agency: QST).
This work was supported by the Engineering and Physical Sciences Research Council (EPSRC) under grant number EP/W032643/1.

\bibliographystyle{IEEEtran}
\bibliography{bib/tools,bib/vqc,bib/qml_examples,bib/quantum_fl, bib/ml_examples,bib/classical_fl,references,bib/fwp,bib/qt,bib/qas,bib/delayed_QC}

\end{document}